# Vision Transformers for Estimating Irradiance Using Data Scarce Sky Images

David Hamlyn [a,*], Sunny Chaudhary [a], Tasmiat Rahman [a]

[a] *School of Electronics and Computer Science, University of Southampton, University Road, Southampton, SO17 1BJ, United Kingdom*

## Abstract

Accurate estimation of diffuse horizontal irradiance (DHI) is critical for optimising photovoltaic system performance and energy forecasting yet remains challenging in regions lacking comprehensive ground-based instrumentation. Recent advancements using Vision Transformers (ViTs) trained on extensive sky image datasets have shown promise in replacing costly irradiance measurement equipment, but the scarcity of long-term, high-quality sky imagery significantly restricts practical implementation. Addressing this critical gap, this study proposes a novel dual-framework approach designed for data-scarce scenarios. First, calculated atmospheric parameters, including extraterrestrial irradiance and cyclic time encodings, are integrated to represent sky conditions without utilising any instrumentation. Next, a sequential pipeline initially predicts synthetic global horizontal irradiance (GHI) and uses it as a feature, to refine DHI estimation. Finally, a dual-parallel architecture simultaneously processes raw and overlay-enhanced fisheye sky images. Overlays are generated through unsupervised, physics-informed cloud segmentation to highlight dynamic sky features. Empirical validation is performed using data from the Chilbolton Observatory, chosen for its temperate climate and frequent cloud variability. To simulate data-scarce conditions, models are trained on a single month (e.g., January) and evaluated across a temporally disjoint, full-year test set. Under this setup, the sequential and dual-parallel frameworks achieve RMSE values within 2-3 W/m² and 1-6 W/m², respectively, of a state-of-the-art ViT trained on the complete dataset. By combining physics-informed modelling with unsupervised segmentation, the proposed method provides a scalable and cost-effective solution for DHI estimation, advancing solar resource assessment in data-constrained environments.

## 1. Introduction

As global solar capacity rapidly scales into the terawatt range, sustaining this growth necessitates highly accurate and reliable irradiance measurements [1]; these are critical for evaluating photovoltaic system performance, forecasting power generation, and ultimately lowering the levelized cost of electricity (LCOE). Accurate measurement of diffuse horizontal irradiance (DHI) is crucial for system design, resource assessment, and energy dispatch; yet continuous high-quality data in this domain remain exceptionally challenging to acquire.

In contrast to global horizontal irradiance (GHI) measurements, which are widely available at global meteorological stations due to the widespread use of Class-A pyranometers, DHI measurements are considerably scarcer globally, primarily because they require specialised and costly sun-tracking instruments [2]. This scarcity severely limits accurate solar forecasting and photovoltaic resource assessment, particularly in remote or developing regions. Conventional irradiance measurement systems represent the industry benchmark for accuracy. However, their substantial initial costs, ongoing maintenance requirements and frequent calibration demands along with the logistical challenges significantly limit their applicability, especially in under-instrumented, remote, or developing regions [2],[3]. This directly undermines the viability of traditional regression-based models, reinforcing the urgent need for alternative approaches [4],[5].

Recent advances in computer vision have shown that vision transformers (ViTs), trained on large-scale sky image datasets, hold significant promise for replacing expensive irradiance measurement equipment [6]. This reinforces the demand for learning architectures capable of producing high-quality predictions from constrained atmospheric or visual input, a challenge our model directly addresses [6].

∗ Corresponding author. Email: dh13g23@soton.ac.uk

To address these limitations, this study builds upon recent advances in ViT architectures for solar irradiance estimation. ViTs have shown considerable success in extracting spatially and radiometrically relevant features from sky imagery. However, their effectiveness is fundamentally constrained by their high parameter count and reliance on self-attention mechanisms, which demand extensive training data to learn feature representations and avoid overfitting. This requirement has been well established in the literature, with studies demonstrating that transformer-based models scale predictably with dataset size, and that insufficient data severely limits their generalisation capabilities [7],[8]. These limitations underscore the importance of developing data-efficient adaptations tailored for sparse-image and low-instrumentation settings.

In contrast, the approach presented herein is explicitly designed for data-constrained environments. It integrates fisheye sky images enhanced via overlay-based segmentation with auxiliary atmospheric parameters, namely, extra-terrestrial irradiance (ETI) and cyclic time encodings, forming a hybrid input representation. Aiming to maximise predictive performance where extensive meteorological datasets or dense image sequences are unavailable.

The proposed framework is validated using measured data from the Chilbolton Facility for Atmospheric and Radio Research (CFARR), Hampshire, UK (51.1445°N, 1.4270°W) [9], [10], [11]. The temperate climate of the site, characterised by frequent and rapid variations in cloud cover, provides a rigorous test environment to evaluate the generality and robustness of the model under realistic and challenging conditions [6].

This paper is structured as follows: First, previous methods for irradiance estimation are reviewed, emphasising upon conventional approaches and recent advancements. The proposed hybrid framework first incorporates calculated atmospheric parameters, including ETI and cyclic time encodings, as input features to improve estimation without relying on direct ground-based measurements. In the sequential pipeline, two models operate in series: the first predicts synthetic GHI, which is then used as an additional feature by a second model to refine DHI estimation. In the parallel dual pipeline, two branches process raw and overlay-enhanced fisheye sky images simultaneously, capturing complementary cloud features to improve prediction accuracy. The models are then validated using data from CFARR. The paper concludes with a discussion of results, comparative analysis, and the practical implications of the proposed method. A key gap in current literature is the absence of a methodology capable of high-accuracy diffuse irradiance estimation using limited sky imagery datasets and minimal instrumentation. Ultimately, the framework introduces a solution for accurate solar estimation without dense data, significantly broadening global accessibility to photovoltaic estimation.

**Contributions and Motivations:** Building on state-of-the-art vision transformers, we introduce a novel hybrid framework to enhance irradiance estimation, reducing reliance on direct measurements and maximising predictive accuracy in data-limited scenarios. Key contributions include: (1) a sequential architecture that first estimates synthetic GHI using one model before integrating it as a feature in a second model, demonstrating a structured approach to improving DHI estimation under diverse atmospheric conditions; and (2) a parallel dual framework that simultaneously processes raw and augmented sky images, capturing complementary information to enhance accuracy in data-constrained scenarios. These innovations advance the field toward a scalable solution for irradiance prediction, as confirmed through experimental validation across varying levels of data availability.

## 2. Data Scarcity Relevance

Solar-irradiance and PV-power forecasting studies consistently report that a full calendar year of historical data is the minimum needed to capture the seasonal cycle and its weather-regime shift [12], [13]. Reflecting this, recent hybrid-modelling reviews recommend machine learning (ML) only if more than one year of data is available, additionally transfer-learning papers show that accuracy improvements flatten once the training window reaches twelve months [14], [15]. Against that baseline, our experiment is deliberately stringent: we train a ViT, a model class acknowledged to be data-hungry and prone to over-fitting when pre-training data are limited, on just one month of 5-min measurements (~2 k usable samples, ≈ 8 % of a year). Other data scarce precedents are (i) Cheng et al. implemented two months of minute-resolution GHI (≈ 50k records $h^{-1}$) labelled their setup “small-data” [16], and (ii) Jeon & Kim whose reinforcement-learning model generalised across a synthetic typical meteorological year (TMY), after points collected over two weeks, a feat enabled by auxiliary forecast inputs and a smoothed test set [17]. By contrast this research, employs no exogenous weather forecasts and is evaluated on an untouched operational year, making its temporal coverage, absolute sample count, and model training collectively leaner than earlier studies. We therefore define data scarcity for transformer-based solar applications as a training corpus that (a) spans ≤ 10 % of an annual cycle or (b) contains < 10 k effective samples; by this operational criterion, our one-month (~2 k-sample) split represents a data scarce benchmark.

## 3. Related Work

Early methods for estimating DHI relied on direct instrumentation or physics-based models. Ground-based instruments, such as pyranometers and sun tracking system with shading balls, deliver reliable measurements; however, their high cost, intensive maintenance, and limited scalability have motivated a shift toward methods that replace physical sensors with calculated atmospheric parameters, synthetic predictions, or sky images [3],[18]. Similarly, physics-informed and empirical models, such as the Fast All-sky Radiation Model and REST2, estimate irradiance using satellite-derived cloud properties and radiative transfer simulations [19]. However, these approaches require extensive atmospheric inputs and exhibit inconsistent performance under heterogeneous or partly cloudy conditions. Simpler empirical relationships based on the clearness index have also been proposed, yet their accuracy diminishes under non-standard sky conditions. Prior to deep learning, studies employing handcrafted features (e.g. cloud fraction and principal component analysis) followed by regression reported mean absolute error (MAE) percentages close to 21.9 %, which motivated the adoption of data-driven methods [18].

The advent of convolutional neural networks (CNNs) marked a significant improvement in irradiance modelling by automatically extracting hierarchical image features from raw data. For instance, Jiang (2020) demonstrated that CNN-based approaches could estimate GHI with a root mean square error (RMSE) of approximately 8.7 %, outperforming conventional baselines such as persistence and ARIMA, which exhibited errors near 29 % and 32 %, respectively [18]. Building on this, Zuo et al. proposed a lightweight CNN image encoder with a fully connected embedding of recent GHI values; the resulting CNN-LSTM cut 10-min-ahead forecast error by 18 % relative to image-only CNNs, underscoring the value of multi-modal conditioning [20]. Subsequent studies, including those from the SolarNet project and research by Salamalikis [21], extended this approach to short-term irradiance forecasting, achieving strong correlation with ground truth measurements, reporting a coefficient of determination of 0.74 for DHI using wide-angle sky imagery [21]. To move beyond 2-D cloud masks, Zuo et al. introduced a luminance-derived proxy for cloud thickness (CT) showing strong correlation with cloud radiative effect, thereby reducing minute-level irradiance uncertainty [22]. Li et al. approached the problem from the segmentation side: their Hybrid Cloud Detection with Intelligent Scene Recognition selects a notable thresholding strategy for each image class and reaches 95.6 % pixel-wise accuracy [23]. Finally, Martinez-Lopez *et al.* partitioned fisheye images into concentric sky regions to isolate solar-proximal clouds, with the overlay strategy which the present study draws upon [24]. Collectively, these studies demonstrate that rich cloud features embedded in all-sky imagery carry substantial predictive power. Nevertheless, CNNs remain constrained by fixed receptive fields that limit their ability to capture wide-field contextual dependencies—an essential factor in the analysis of hemispherical sky images.

To address these limitations, ViTs have been adopted due to their self-attention mechanisms, which facilitate the capture of long-range spatial dependencies across entire images. These self-attention mechanisms were originally introduced in 'Attention is All You Need' [25] and later adapted for computer vision tasks [7]. Mercier et al. (2024) demonstrated that data-efficient image transformers (DeiTs) could estimate GHI and DHI from all-sky imagery with RMSE of approximately 52 W/m² and 31 W/m², respectively, without auxiliary meteorological features [6]. Their work showed that ViTs better identified cloud patterns near the solar disc compared to CNNs. In another study, Mercier (2023) introduced a temporal encoder-decoder model—termed the Solar Irradiance Anticipative Transformer—that forecasts irradiance 15 minutes ahead using sequences of all-sky images, outperforming smart persistence benchmarks by over 20 % in skill score [26].

Building on these innovations, recent investigations have integrated multimodal data to enhance irradiance prediction. Liu et al. (2023) proposed a transformer-based multimodal framework that fuses optical flow derived from ground-based sky images with historical irradiance time series via cross-modal attention. Their model achieved a normalised RMSE of 4.28 % for 10-minute-ahead global irradiance forecasting [27]. Other studies have applied transformer-based or hybrid attention architectures to satellite data. For instance, Li et al. (2025) introduced the SolarFormer model, which uses geostationary satellite observations and a gated recurrent unit (GRU) to estimate global irradiance with lead times up to three hours [28]. Chen et al. (2023) developed a convolutional-attention network that selectively processes multispectral satellite bands, achieving a 20.57 % normalised error—exceeding the performance of the National Solar Radiation Database (NSRD) by more than 5 % [29]. In contrast to Mercier et al. (2024) [6], which solely relies on raw sky imagery, the present study uniquely integrates synthetic GHI, overlay-based segmentation, and physics-informed auxiliary features to achieve higher accuracy under sparse-data conditions.

Despite their successes, existing transformer-based models typically assume access to dense, long-term datasets or high-resolution imagery, severely constraining their applicability in data-scarce or remote regions. To overcome this fundamental barrier, this study innovatively integrates synthetic GHI predictions, cyclic time encodings, and lightweight, unsupervised cloud segmentation methods within a transformer architecture, explicitly designed for sparse data environments. Moreover, the estimation of DHI has received less attention compared to global components, and cloud segmentation, particularly lightweight unsupervised methods based on threshold logic, is rarely integrated within transformer architectures. To address these issues, the present study proposes a dual framework that combines synthetic GHI predictions, cyclic time encodings, and overlay-enhanced fisheye imagery. By merging physics-informed atmospheric features with a cloud-focused segmentation strategy in a ViT framework, this work improves the reliability and applicability of diffuse irradiance estimation in data-scarce environments.

# 4. Proposed Framework

Throughout the frameworks described below, sky images are first filtered to remove non-informative borders and camera obstructions. A black mask is applied to every non-sky pixel, and the remaining non-zero pixels constitute the binary validity mask $M_{valid}$, ensuring that all subsequent modules operate exclusively on informative sky content.

## *4.1. Incorporating Cyclic Time Encodings and Extra-Terrestrial Irradiance*

The integration of ETI and cyclic time encodings enhances the accuracy of ViT models by providing essential temporal and atmospheric context [30], [31], [32]. Cyclic time encodings help the model understand that the input images are not random snapshots, but part of a continuous temporal sequence, capturing the natural progression of time such as one day following another, along with recurring patterns in solar energy availability due to the rotation of the Earth and seasonal cycles. ETI, in turn, is a deterministic upper bound on incident irradiance, calculated from the solar constant, the Earth–Sun distance correction factor, and the solar zenith angle [24]. ETI encapsulates fundamental atmospheric and solar position parameters, offering a physics-informed feature that aids in predicting DHI [30], [32].

## *4.2. Sequential Stream Framework Utilising Synthetic GHI*

To capture the improvement measured GHI adds in refining DHI estimation [33], the sequential pipeline incorporates synthetic GHI predictions from a preliminary model: the first stream produces the synthetic GHI, and the second stream combines this value with the image to predict DHI. The synthetic input provides additional context, helping the second stream remain accurate under variable atmospheric conditions even when irradiance data are sparse or missing.

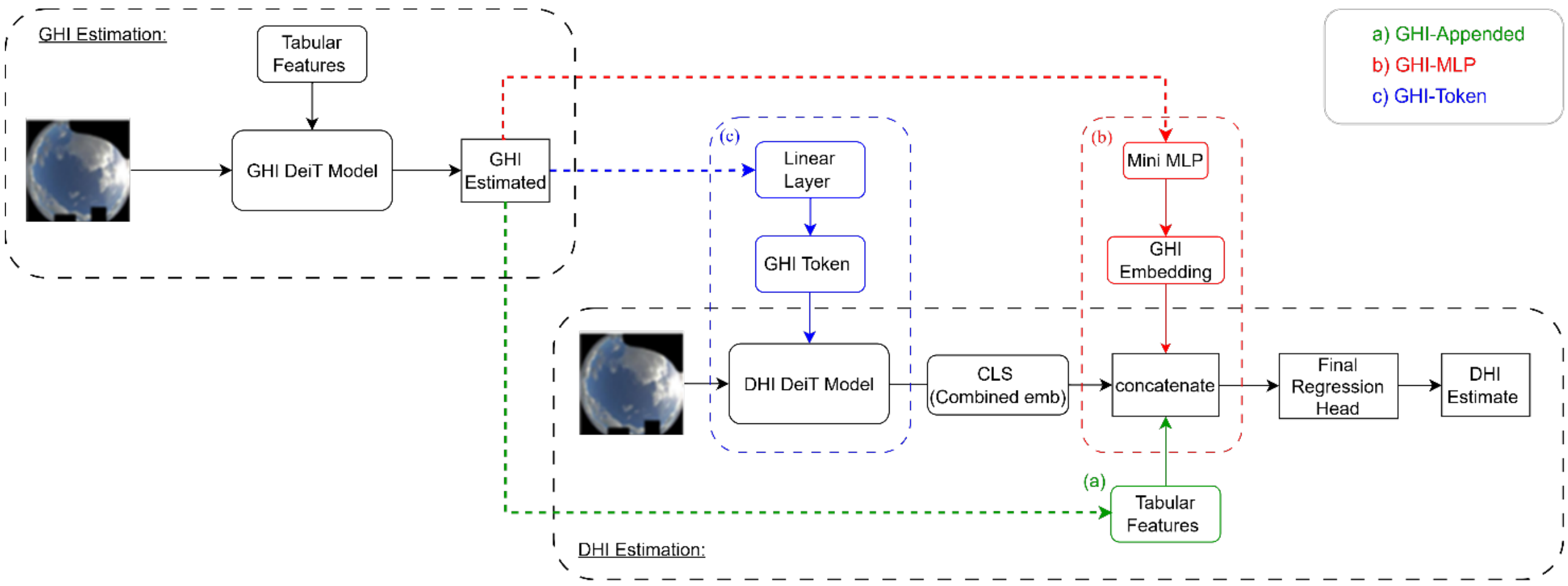

Figure 1 (a) – High-level schematic of the two-stage Sequential framework: Stage 1 predicts synthetic GHI from the fisheye image; Stage 2 ingests the same image plus this synthetic GHI to regress DHI.

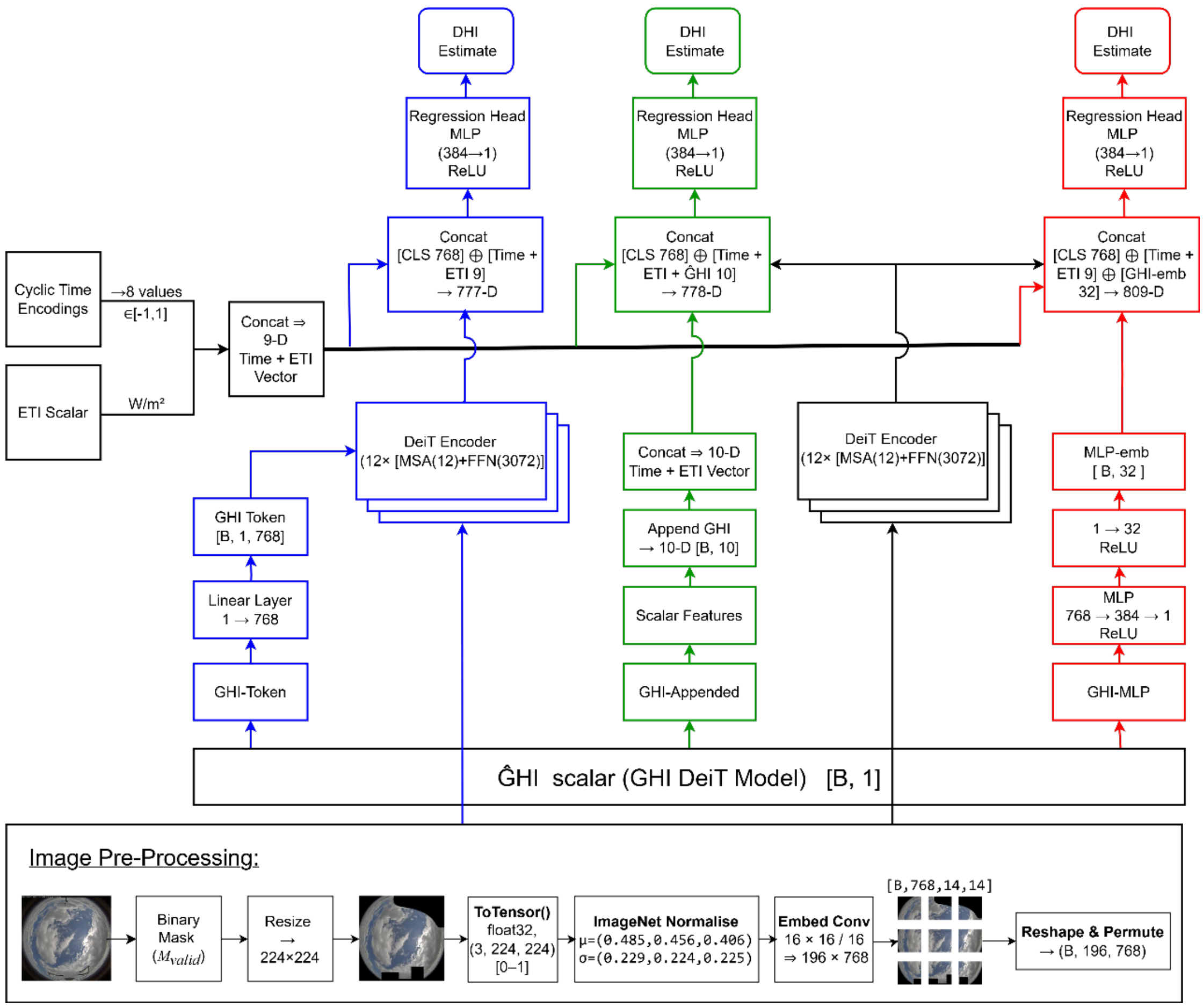

Figure 1 (b) – Detailed overview of the sequential framework showing preprocessing, synthetic-GHI feedback, and the three colour-coded injection strategies (GHI-Appended = green, GHI-MLP = red, GHI-Token = blue).

Figure 1 (a) provides a schematic overview, while Figure 1 (b) unpacks each component in detail. Starting with a fisheye frame, we (i) mask non-sky pixels, (ii) resize and normalise colour channels, (iii) split the image into 16 × 16 patches, and (iv) prepend a CLS token plus cyclic time and ETI scalars. All variants share a common backbone: a DeiT encoder that transforms each 224 × 224 sky image into a 768-D class token, followed by our regression head—two fully connected layers (Linear → Rectified-Linear-Unit (ReLU) → Linear)—which outputs the DHI in W/m². The first stage delivers a single, scalar number, the synthetic GHI. What follows is an ablation in which that scalar is feed back into the second transformer at three progressive depths. Although every variant starts from identical inputs—196 image tokens, a [CLS] token, and a 9-D tabular vector (eight cyclical-time encodings plus ETI)—they differ in how and where the scalar is introduced, whether it is embedded, transformed, or left raw, and how much it influences downstream computations.

GHI-Appended (green): This minimal baseline appends the raw GHI scalar directly to the 9-D tabular vector, producing a 10-D input. The scalar bypasses the transformer and interacts with image content only in the final regression head. It introduces no new parameters and provides the narrowest path for GHI to affect the prediction.

GHI-MLP (red): In the following variant we first project the scalar through a lightweight multi-layer perceptron (MLP: two or more stacked linear layers with a non-linearity; here Linear 1 → 32 → ReLU and obtain a 32-D embedding. After the DeiT encoder has distilled the image into its [CLS] token, we concatenate this 32-D vector with the 9 tabular features, forming an 809-D input to the regression head. GHI-MLP allows modest non-linear transformation of GHI but does not expose it to self-attention.

GHI-Token (blue): Here the 1-D GHI scalar passes through a linear layer (dense layer) that expands it to the DeiT hidden size (1 → 768). The resulting vector is inserted as a new token directly after the [CLS] and distillation tokens in the DeiT embedding stack (torch.cat along the sequence axis). Because this GHI-Token joins the sequence before the first multi-head self-attention layer, it participates in all 12 encoder blocks; consequently, the tokens can attend to absolute irradiance and vice-versa. Empirically, this full-depth interaction yields the lowest RMSE but adds an extra token to the attention matrix (see Table 4).

These three mutually exclusive strategies for introducing the synthetic-GHI into the sequential pipeline offer progressively deeper integration. GHI-Appended (green) provides minimal gains; Table 4 demonstrates this modest reduction in the 1-month RMSE (43.3 versus 42.2 W/m² for the ETI + Time baseline). GHI-MLP (red) although providing a late importance cue again falls short of the expected benefit. GHI-Token (blue) which projects the scalar to an embedding and inserts it as an additional token at the encoder input, allowing every self-attention layer to condition on irradiance context; this design consistently attains the lowest errors across all data regimes (e.g., 33.8 W/m² at 1 month, 33.1 W/m² at 3 months and 24.6 W/m² with full-year training), and is therefore adopted as the default configuration for the sequential framework.

## *4.3. Parallel Dual Approach*

The parallel dual-stream approach processes two complementary representations of fisheye sky images: the original raw images and overlay-enhanced images generated via hybrid segmentation. The overlay-enhancement involves combining threshold-based cloud masks and super-pixel refinements to effectively highlight cloud dynamics.

### *4.3.1. Overlay Generation Using Hybrid Thresholding and Super Pixel Segmentation*

To enhance cloud-detection accuracy we adopt the hybrid strategy in [24], which applies a fixed threshold on the normalised blue–red ratio (nBRR) when the sun is obscured [34] and, when the sun is clear, refines the mask with Simple Linear Iterative Clustering (SLIC) super-pixels and local Otsu thresholds as described in [35]. This methodology focuses on computational efficiency and adaptability to diverse atmospheric conditions. Using the sky image already masked by $M_{valid}$, the framework first determines whether the sun is visible or obscured by clouds. If the sun is blocked, intensity-based thresholding is used for mask generation. The solar azimuth and altitude are mapped onto the fisheye image using a projection model. The zenith angle is calculated from the solar altitude, and the position of the sun is projected onto the image plane using standard radial and angular transformations, as described in [36].

Subsequent processing focuses exclusively on the cropped region containing the solar disc. The sun is classified as clearly visible if more than 95 % of pixels within this cropped area exceed a pixel intensity threshold of 210. This threshold was empirically determined through rigorous sensitivity analysis against ground truth observations, ensuring reliable discrimination between clear-sky and cloud-obstructed sun conditions across diverse atmospheric states.

### *4.3.2. Threshold-Based Cloud Mask Generation*

Cloud regions were initially identified using a greyscale intensity threshold. The greyscale image, $I_{gray}(x,y)$, was normalised to the range $[0,1]$ as: $I_{graynorm}(x,y) = I_{gray}(x,y)/255$. The cloud mask, $M_{cloud}$, was generated by applying a predefined threshold, $T_{optimal}$:

$$M_{cloud}(x,y) = \begin{cases} 1, & I_{graynorm}(x,y) > T_{optimal} \\ 0, & otherwise \end{cases} \quad (1)$$

The optimal threshold, $T_{optimal} = 0.3946$ was determined by sweeping candidate values and selecting the one whose cloud masks most closely tracked two sky-state metrics: (i) the diffuse fraction $K_d = DHI/GHI$ and (ii) the cloud ratio (CR), the fraction of pixels classified as cloud among all valid-sky pixels; see Equation 3. A fixed threshold on the nBRR (> 0.6), widely used in the earlier studies, produced a weaker $CR - K_d$ correlation (R=0.68 versus R=0.78 for the greyscale mask); the greyscale threshold was therefore adopted. Because this tuning criterion links the mask directly to radiative physics rather than to hand-labelled data, it minimises false detections across a broad range of atmospheric conditions.

### *4.3.3. Super Pixel Segmentation and Adaptive Thresholding*

To refine the cloud mask and improve detection accuracy under heterogeneous sky conditions, we employed a combination of SLIC super pixels and Otsu's adaptive thresholding [35]. The red-blue intensity difference feature, $\Delta_{RB}(x,y)$, was computed as: $\Delta_{RB}(x,y) = I_{red}(x,y) - I_{blue}(x,y)$ where $I_{red}$ and $I_{blue}$ denote the red and blue channels of the RGB image, respectively. The normalised feature, $\Delta_{RBnorm}$, was calculated as:

$$\Delta_{RBnorm}(x,y) = \left(\frac{\Delta_{RB}(x,y) - min(\Delta_{RB})}{max(\Delta_{RB}) - min(\Delta_{RB})}\right) \times 255 \quad (2)$$

Super pixels were generated by partitioning the image into 100 regions, an empirically chosen balance between spatial detail and runtime, using the SLIC algorithm. Otsu's method was applied to each super pixel to calculate adaptive thresholds, $T_{local}$, based on the distribution of $\Delta_{RBnorm}$ within the region: $T_{local} = (0.5 \times T_{Otsu}) + (0.5 \times T_{global})$, where $T_{Otsu}$ represents the threshold computed for the specific super pixel and $T_{global}$ is a global threshold for the entire image. This refinement process ensures accurate cloud detection even in the presence of localised brightness variations or partial occlusion.

### *4.3.4. Overlay Generation and Cloud Ratio Calculation*

To visualise the detected cloud regions, the binary mask $M_{cloud}$ was overlaid on the original image using a semi-transparent colour layer. The blending operation is defined as: $I_{overlay}(x,y) = \alpha \times I_{RGB}(x,y) + (1-\alpha) \times C_{mask}$, where alpha is the transparency factor, and $C_{mask}$ is the mask colour (e.g., red for cloud regions). The CR, was calculated as:

$$CR = \frac{\sum M_{cloud}(x,y)}{\sum M_{valid}(x,y)} \quad (3)$$

The described overlay generation method was implemented using Python and integrated into the overall framework. This approach demonstrated consistent cloud segmentation performance across diverse sky conditions, laying a strong foundation for accurate irradiance modelling as demonstrated by the CR-$K_d$ correlation. The segmentation identifies cloud pixels via classifier free adaptation of the nBRR and hybrid thresholding methods [24]. Within the current framework, the overlay mask functions solely as an auxiliary signal to support irradiance estimation, rather than serving as an independent optimisation objective. High-precision cloud masks tailored for segmentation remain available; for instance, the scene-aware hybrid approach by Li et al. [23] achieves approximately 95 % pixel-level accuracy without requiring dense per-pixel supervision. In contrast, fully supervised architectures such as U-Net and DeepLab-v3+ typically report Intersection-over-Union (IoU) scores between 0.7 and 0.92 [37]. However, the enhanced accuracy offered by supervised models comes at the cost of extensive manual annotation and increased computational load. Notably, our heuristic mask improves DHI RMSE by 6.8 W/m² relative to the Normal-DeiT ablation (Table 5), demonstrating sufficient physical consistency for effective downstream use. This overall segmentation approach can be seen in Figure 2. Further details of the threshold logic and sun-blocking considerations mirror earlier work [24] but are adapted to remain accurate under unlabelled scenarios. The adaptations as described above ensure continued accuracy across unlabelled datasets, a necessity for deployment where supervised data is scarce or unavailable.

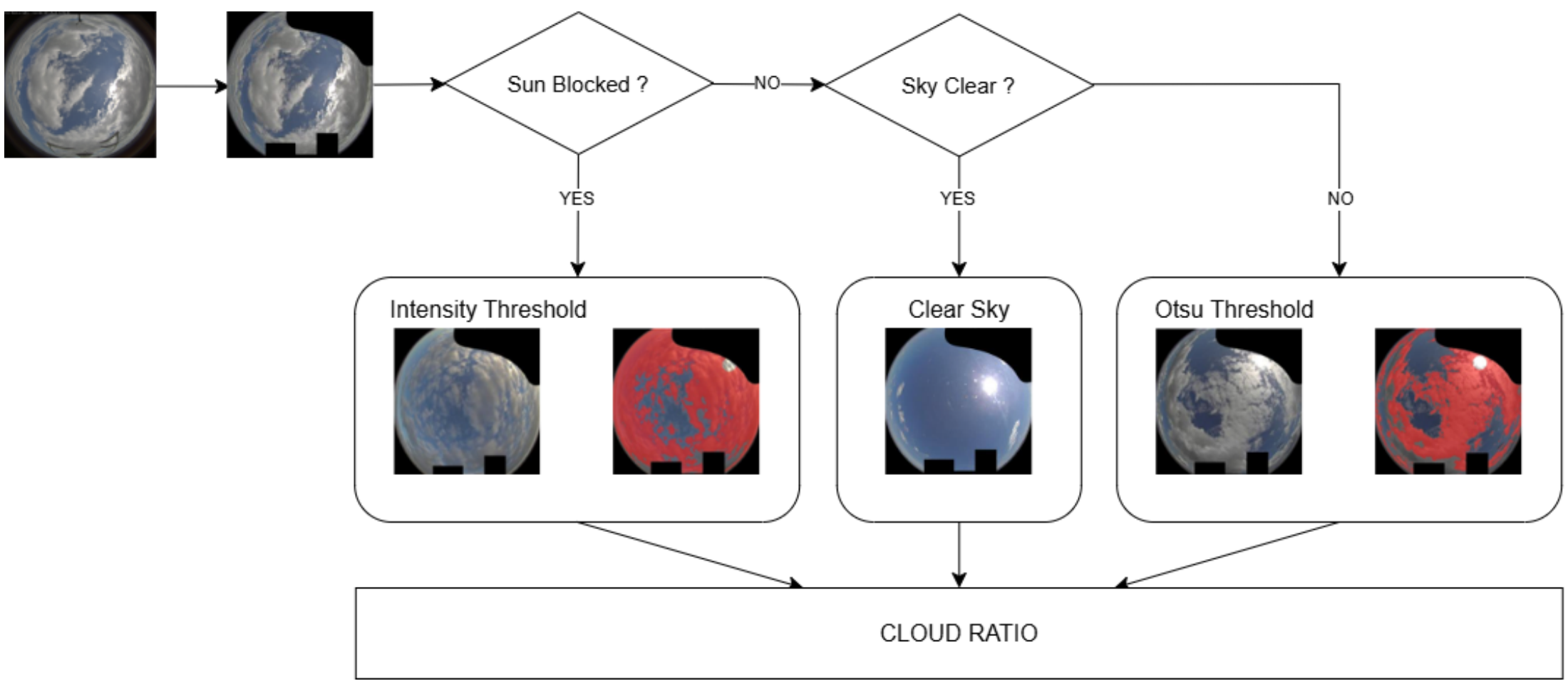

Figure 2 – Illustration of the overlay process for sky image segmentation and synthetic enhancement. Though not a classifier, this method reliably highlights cloud patterns to improve DHI estimation.

*4.4. Parallel Dual Structure for Cloud Segmentation and Feature Inclusion*

In the proposed parallel dual framework Figure 3, two DeiTs process different representations of sky images in parallel. The first using the overlay processed image as described in 4.3.4, whilst the other branch processes the raw sky image.

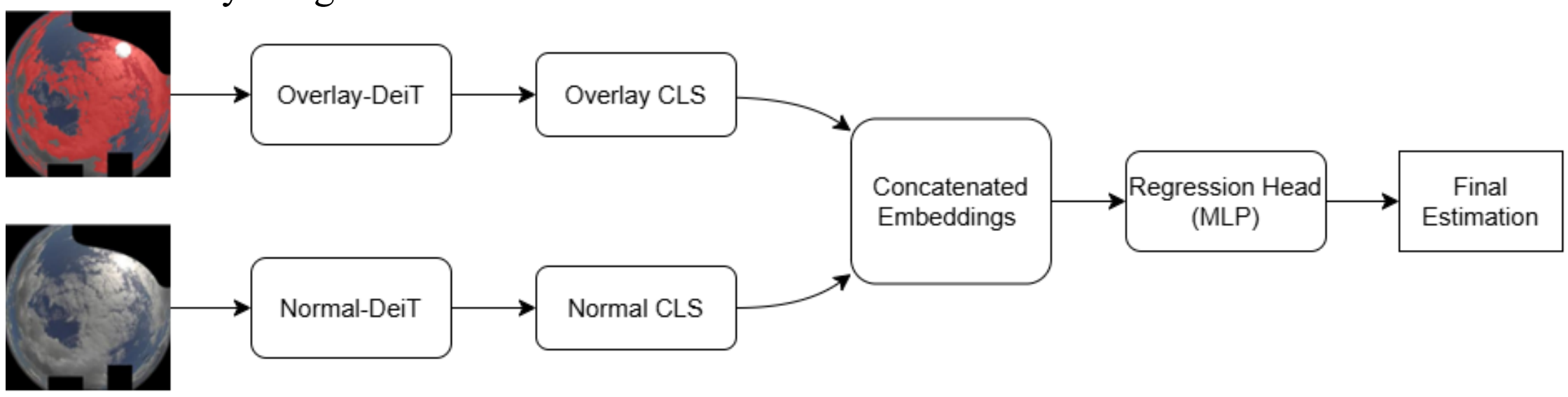

Figure 3 – Preprocessing pipeline for sky images, involving solar position estimation and segmentation of clouds via thresholding. Pre-segmented images are fed into the Overlay-DeiT branch, whilst raw images are processed by Normal-DeiT branch. Model embeddings are concatenated with tabular features (e.g., ETI, time) for final DHI estimation.

Each DeiT encoder produces an embedding that is concatenated with meteorological features, including ETI and cyclical time encodings. The resulting vector is passed to a two-layer MLP regression head to estimate DHI. This dual-stream structure mitigates dependence on detailed cloud labels by leveraging threshold-based overlays. To ensure that performance improvements are not attributable solely to increased parameter count, two additional variants were introduced: (i) a single-stream DeiT with concatenated normal and overlay images as a six-channel input, with all architectural dimensions (hidden size, number of attention heads, and feedforward width) doubled to match the dual-stream capacity, and (ii) separate single-stream DeiTs using either normal or overlay inputs with the alternate modality zeroed out. These ablations confirm that gains arise from architectural design rather than model scale.

## 5. Dataset

Data was collected at the CFARR using a fisheye lens sky camera at 5-minute intervals. All-sky images were acquired with the "cloud-camera 2" system, which incorporates a factory-calibrated 180° fisheye optic. The JPEG frames are archived as produced by the on-board firmware; no additional geometric rectification is performed by the observatory [11]. In our workflow, raw fisheye projection was retained to avoid unnecessary preprocessing, model-specific re-projection, and the risk of error amplification. The models therefore learn directly from the native angle projection; no "unfolding" or remapping was applied. Calibrated secondary-standard pyranometers provided ground-truth GHI and DHI. In certain sampling periods, DHI data exhibited gaps; thus, a secondary objective of this work is to validate a flexible, segmentation-based approach for images and a sequential pipeline for bridging limited data. To ensure data quality, images captured between 11 p.m. and 3 a.m. were excluded, and all negative values of DHI and GHI were filtered out. Sampling intervals with missing DHI readings were omitted, and the study window was shifted to the nearest contiguous period with continuous measurements, to ensure no imputation was required. Each image was matched directly to its corresponding irradiance value based on exact timestamp alignment. Sky images were resized to $224 \times 224$ pixels and standardised via ImageNet preprocessing [38]. Data augmentation (random rotations up to 45°) was applied to strengthen generalisation. In this study we adopt a deliberately minimalist augmentation scheme: aside from ImageNet normalisation and the rotation, no colour perturbations, blurring, or similar transforms are applied. This keeps architectural comparison transparent. Investigating richer, sky-specific augmentations remains valuable future work. A binary mask was applied to each image to exclude non-sky regions and camera artefacts, retaining only valid sky pixels for model input as per 4.3.1.

The Folsom dataset [39] offers three years (2014 – 2016) of 1-min all-sky imagery and co-located irradiance data gathered in Folsom, California (38.642 °N, 121.148 °W). Each daytime frame, captured every 60 s by a wide-angle sky camera, is timestamp-paired with ground-truth GHI and DHI from secondary-standard pyranometers. The release also bundles concurrent weather, satellite and NWP variables, making it a common benchmark for cross-site solar-forecasting research. For comparability, we apply the same preprocessing, to allow for evaluation on a distinct site without site-specific alterations [39].

To evaluate performance under varying data availability conditions, training was conducted over multiple temporal subsets ranging from 1 to 12 months, beginning in May 2016 and extending through April 2017.

Within each subset, 80 % of the samples were used for training and the remaining 20 %, selected at random from the same period, served as an internal validation set. Each subset is always tested on a full year from August 2017 to August 2018, ensuring no temporal overlap with training and maintaining a consistent, seasonally representative full-year evaluation window. This design simulates realistic data-scarce scenarios whilst preserving fair cross-period comparisons and enabling reproducible assessments of generalisation performance.

# 6. Implementation

The backbone model employs a ViT (pretrained on ImageNet) [40]. Images are split into $16 \times 16$ patches, then passed through multi-head self-attention layers. Initially, an Image-Only DeiT replicating recent literature processes the images without auxiliary features [6]. The architecture is then extended with ETI and cyclic time encodings to create the Baseline (ETI + Time) model. Next, in the sequential approach, a single DeiT model first estimates GHI; this estimate is then integrated into the DHI prediction pipeline via three strategies: (1) as an appended tabular feature (GHI-Appended), (2) as an independent pre-regression input passed through a small MLP (GHI-MLP), or (3) as an injected token within the transformer encoder (GHI-Token), each designed to enhance spatiotemporal feature capture. Amongst these, early-stage token integration achieved the highest accuracy, benefiting from attention-based learning that captures both spatial and temporal context effectively. Finally, in the parallel setup, the first DeiT ingests the threshold-segmented overlay sky images, whilst the second DeiT uses unaltered sky images. The resulting embeddings are fused with standard meteorological features including ETI [30] and cyclic time encodings. DHI prediction emerges from a MLP regression head. Tables 1-3 presents the detail of the layer structures, parameter counts, and key hyper-parameters for transparent replication of results.

Table 1 – Architecture-summary table

| Variant | Backbone | Encoder blocks | Hidden dim | MLP dim | Heads | Param ($\boldsymbol{M^+}$) |
|---|---|---|---|---|---|---|
| Single Stream Variants | 1 × DeiT | 12 | 768 | 3072 | 12 | ≈ 86 |
| Sequential Stream | (1+1) DeiT | 12+12 | 768 | 3072 | 12+12 | ≈ 86 (+86) |
| Dual Stream | 2 × DeiT | 12+12 | 768 | 3072 | 12+12 | ≈ 172 |
| Concatenation -Normalised | 1 × DeiT | 24 | 768 | 3072 | 12 | ≈ 172 |
| Concatenation -Stream | 1 × DeiT | 12 | 2304 | 3072 | 24 | ≈ 436 |

Table 2 – Hyper Parameter Table: Global Default and Search Space

| Hyper-parameter | Default value | Grid / tuned range |
|---|---|---|
| Optimiser | AdamW | — |
| Epochs | 10 | {10, 20} |
| Batch size | 16 | {8, 16, 32} |
| Learning-rate (η) | $1 \times 10^{-4}$ | log-uniform [$1 \times 10^{-5}$ – $3 \times 10^{-4}$] |
| Weight-decay (w) | $1 \times 10^{-6}$ | {$1 \times 10^{-8}$, $1 \times 10^{-6}$, $1 \times 10^{-4}$} |
| Reg-head hidden dim | 384 | {256, 384, 512} |
| GHI-MLP embed dim | 32 | {16, 32, 64} |
| | *Concatenation-Stream Multipliers* | |
| Hidden size ($\mathbf{M_d}$) | 3 | {2, 3, 4} |
| Heads ($\mathbf{M_h}$) | 2 | {1, 2} |
| FFN ($\mathbf{M_{ffn}}$) | 1 | {1, 2} |

Table 3 – Hyper Parameter Table: Variant-specific overrides

| Variant | $\boldsymbol{\eta}$ | $\boldsymbol{w}$ | $\boldsymbol{M_d}$ | $\boldsymbol{M_h}$ | $\boldsymbol{M_{ffn}}$ |
|---|---|---|---|---|---|
| Image-Only (DeiT) | ✓ | ✓ | — | — | — |
| Baseline (ETI+Time) | ✓ | ✓ | — | — | — |
| GHI-Appended | ✓ | ✓ | — | — | — |
| GHI-MLP | ✓ | ✓ | — | — | — |
| GHI-Token | ✓ | ✓ | — | — | — |
| Overlay-DeiT (single overlay) | ✓ | $2.5 \times 10^{-6}$ | — | — | — |
| Normal-DeiT (single normal) | ✓ | $2.5 \times 10^{-6}$ | — | — | — |
| Parallel Dual-Stream | ✓ | $2.5 \times 10^{-6}$ | — | — | — |
| Concatenation -Normalised | ✓ | $2.5 \times 10^{-6}$ | — | — | — |
| Concatenation -Stream | $2.61 \times 10^{-5}$ | $1.29 \times 10^{-8}$ | 3 | 2 | 1 |

Training used 10 epochs for all models with a batch size of 16 and the AdamW optimiser with a base learning rate (LR) of $1 \times 10^{-4}$ [41]. Within each architecture family a 50-trial Optuna [42] search (η, w, head width) fixed the hyper-parameters reported in Tables 1–3; these were then held constant in all subsequent runs. For the concatenation stream we additionally trained a parameter-matched Concatenation-Normalised variant (width scaled down) and a Concatenation-Stream variant (width doubled) to disentangle architectural gains from pure capacity effects, ensuring that performance differences arise purely from architectural changes. All experiments were carried out on a workstation with an NVIDIA Quadro RTX 6000 GPU (24 GB VRAM) and an Intel Core i7-7700K CPU (64 GB RAM). Code was implemented in PyTorch [43], with data parallelism for the dual-stream models, and all tensor operations used standard cuBLAS FP32 kernels.

# 7. Results

## *7.1. Performance*

To systematically evaluate model performance under varying levels of data availability, assessments were conducted across scenarios ranging from highly constrained (1-month dataset) to comprehensive (12-month dataset). Figures 4 and 5 explicitly illustrate that the frameworks provide substantial predictive advantages primarily under severely limited data conditions. Specifically, the dual framework compensates effectively for reduced dataset sizes by capturing nuanced atmospheric features through synthetic data integration and cloud segmentation enhancements. However, as data availability increases, the relative benefit of the dual framework decreases, indicating that the framework is most effective specifically in sparse-data scenarios, as expected. Although MAE and $R^2$ follow similar trends (see Table 4 and Table 5), RMSE is presented as the primary metric due to its heightened sensitivity to larger errors and its widespread use in the literature. Training was repeated across five random seeds for all temporal subsets, and the average RMSE standard deviation across all months was within ±1.76 W/m² for the baseline model, ±2.29 W/m² for the GHI-Token model, and ±1.45 W/m² for the parallel dual model, demonstrating model stability.

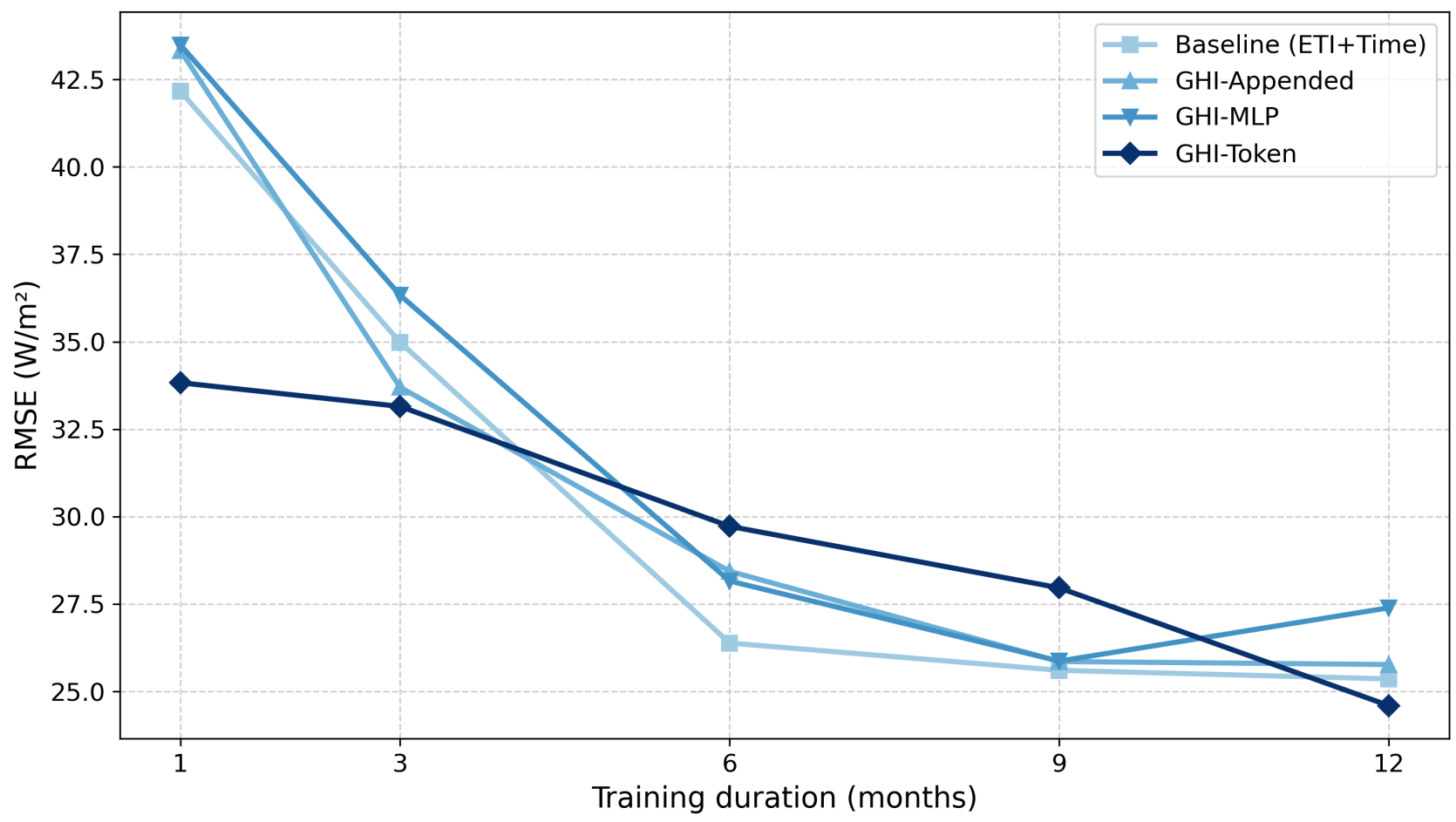

Figure 4 – Performance of Sequential Framework on DHI Estimation

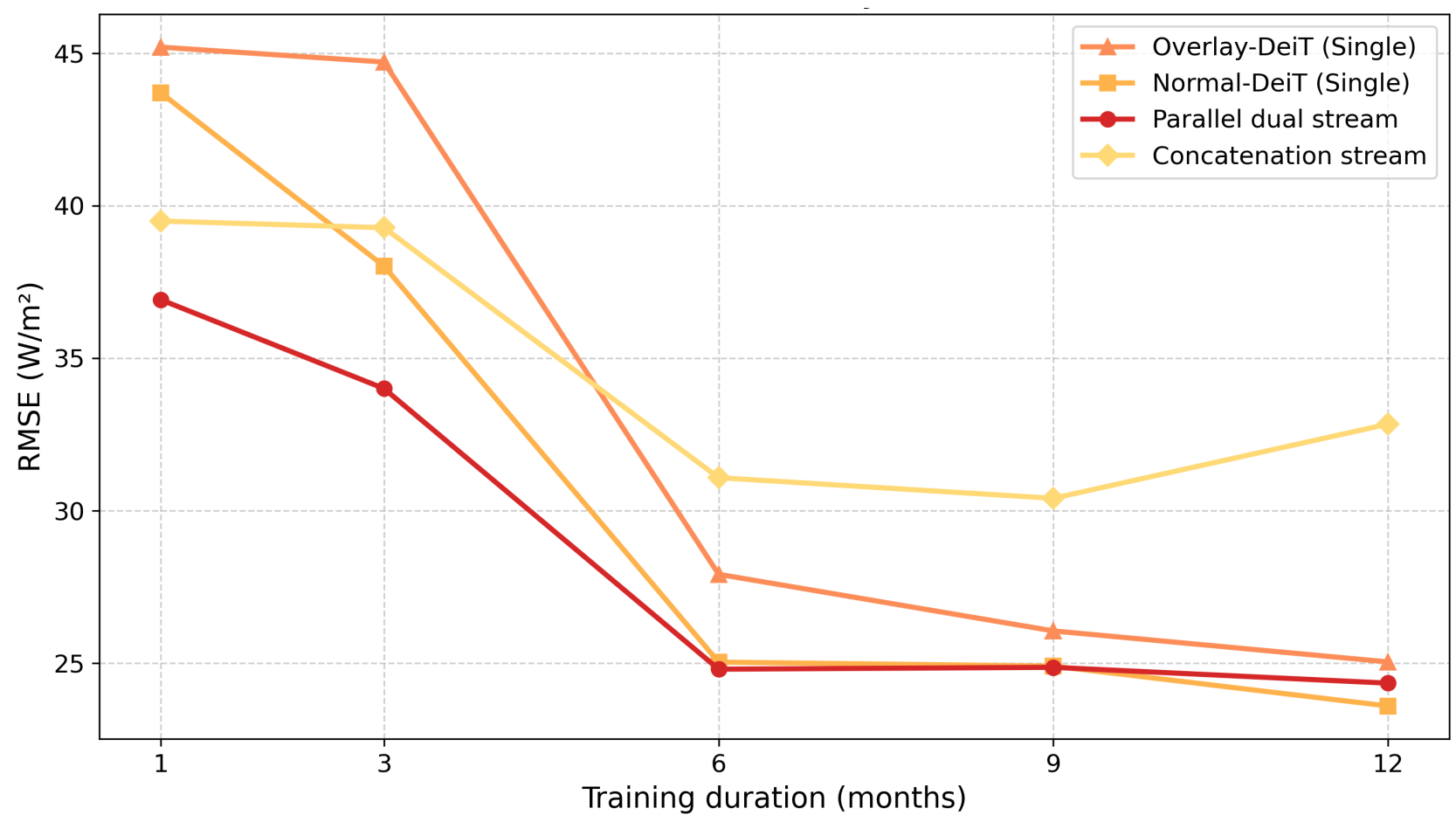

Figure 5 – Impact of Parallel Dual Framework on DHI Estimation

$$RMSE = \sqrt{\frac{1}{n}\sum_{k=1}^{n}(y_k - \widehat{y_k})^2} \quad (4)$$

$$MAE = \frac{1}{n}\sum_{k=1}^{n} |y_k - \widehat{y_k}| \quad (5)$$

$$R^2 = 1 - \frac{\sum_{k=1}^{n}(y_k - \widehat{y_k})^2}{\sum_{k=1}^{n}(y_k - \overline{y_k})^2} \quad (6)$$

### *7.2. Model Benefits in Limited Dataset Scenarios*

When the training set is constrained (e.g., only a subset of months is available), the parallel pipeline outperforms direct DHI modelling. Incorporating the proposed frameworks helps the network contextualise the overall irradiance level of each sky image, partially compensating for the lack of large training volumes. Table 4 showcases how the two-stage approach of the sequential model reduces RMSE by approximately 20 % relative to a single-stage Baseline approach when ~8.3 % (1/12) of the data is used.

Table 4 – Performance Metrics for GHI-Injection Strategies (Sequential Pipeline)

| Time (Months) | Injection Strategy | MAE | RMSE | $R^2$ |
|---|---|---|---|---|
| 1 | Image-Only (no auxiliary features) | 39.405 | 59.488 | 0.717 |
| 1 | Baseline (ETI + Time) | 30.299 | 42.159 | 0.858 |
| 1 | GHI-Appended | 31.201 | 43.328 | 0.850 |
| 1 | GHI-MLP | 31.602 | 43.486 | 0.849 |
| 1 | GHI-Token | **22.366** | **33.827** | **0.908** |
| 3 | Image-Only (no auxiliary features) | 48.198 | 68.480 | 0.625 |
| 3 | Baseline (ETI + Time) | 24.769 | 34.983 | 0.902 |
| 3 | GHI-Appended | 23.842 | 33.696 | 0.909 |
| 3 | GHI-MLP | 25.948 | 36.329 | 0.894 |
| 3 | GHI-Token | **21.804** | **33.149** | **0.912** |
| 12(Full-Year Data) | Image-Only (no auxiliary features) | 25.144 | 41.182 | 0.864 |
| 12(Full-Year Data) | Baseline (ETI + Time) | 15.586 | 25.359 | 0.949 |
| 12(Full-Year Data) | GHI-Appended | 16.241 | 25.770 | 0.947 |
| 12(Full-Year Data) | GHI-MLP | 17.157 | 27.392 | 0.940 |
| 12(Full-Year Data) | GHI-Token | **14.953** | **24.596** | **0.952** |

Table 5 – Performance Comparison of DHI Model Variants (Parallel Pipeline)

| Time (Months) | Framework Variant | MAE | RMSE | $R^2$ |
|---|---|---|---|---|
| 1 | Overlay-DeiT (Single) | 33.278 | 45.194 | 0.790 |
| 1 | Normal-DeiT (Single) | 32.233 | 43.694 | 0.804 |
| 1 | Parallel Dual Stream | **26.594** | **36.916** | **0.861** |
| 1 | Concatenation-Normalised | 33.495 | 45.043 | 0.792 |
| 1 | Concatenation Stream | 27.894 | 39.495 | 0.840 |
| 3 | Overlay-DeiT (Single) | 33.163 | 44.706 | 0.795 |
| 3 | Normal-DeiT (Single) | 28.599 | 38.021 | 0.851 |
| 3 | Parallel Dual Stream | **24.125** | **34.006** | **0.881** |
| 3 | Concatenation-Normalised | 28.399 | 41.633 | 0.822 |
| 3 | Concatenation Stream | 28.748 | 39.276 | 0.841 |
| 12(Full-Year Data) | Overlay-DeiT (Single) | 15.769 | 25.047 | 0.936 |
| 12(Full-Year Data) | Normal-DeiT (Single) | 14.956 | 23.602 | 0.943 |
| 12(Full-Year Data) | Parallel Dual Stream | 15.306 | 24.35 | 0.939 |
| 12(Full-Year Data) | Concatenation-Normalised | 26.512 | 39.069 | 0.843 |
| 12(Full-Year Data) | Concatenation Stream | 21.124 | 32.840 | 0.889 |

Most critically, the dual framework excels when data is sparse, validating its use in deployment scenarios where ground truth is limited or irregular.

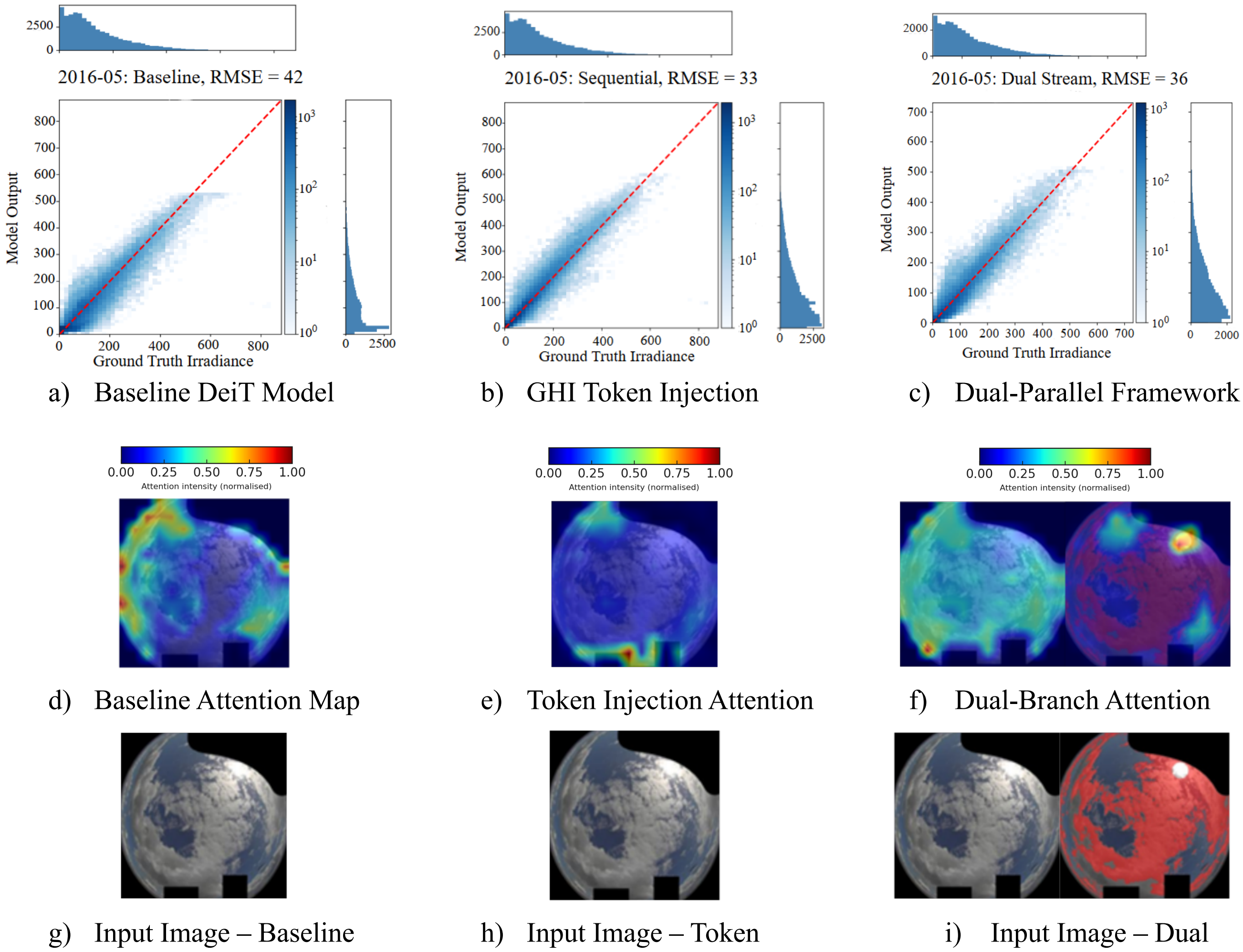

Figure 6 – Prediction accuracy and attention visualisation for three model variants: (a, d, g) baseline DeiT, (b, e, h) DeiT with GHI token injection, and (c, f, i) dual-parallel DeiTs processing raw and overlay-enhanced images. Scatter plots compare predicted and ground truth values. Heatmaps show attention regions, whilst input images reveal segmentation-based enhancement.

In addition to quantitative performance metrics, Figure 6 presents the attention rollout map visualisations [44] in certain instances, it is observed that the dual-parallel framework directs focus to regions that might otherwise be under-emphasised without the overlay enhancement. By effectively combining the strengths of the raw image branch and the overlay branch, the model concentrates on key features, in several samples resembling the attention patterns observed in the baseline method, whilst also capturing complementary contextual cues from the overlay. This balanced focus enables the model to extract critical information from sparse data, thereby improving prediction accuracy in limited-data scenarios.

To explicitly benchmark the effectiveness of the proposed frameworks against existing state-of-the-art results, the performance is compared directly with the recent ViT model from Mercier et al. (2024), as both studies employed the identical Chilbolton dataset under the same climatic conditions. Table 6 summarises this direct comparison, highlighting clear improvements achieved through these approaches.

Table 6 – Comparative performance metrics of the proposed frameworks versus Mercier et al. (2024)

| Method | MAE | RMSE | $R^2$ |
|---|---|---|---|
| Mercier et al. (2024) | 19 | 31 | - |
| Proposed Parallel Dual | 26.6 | 36.9 | 0.86 |
| Proposed Sequential | 22.4 | 33.8 | 0.91 |

Mercier et al. (2024) used the full dataset with intra-month testing (days 5–9) and evaluation (days 15–19), enabling high data density for training. In contrast, this study employs a temporally disjoint validation strategy, training on only one month and testing on a separate full-year period. Despite this stringent setup, the sequential framework achieves an RMSE of 33.8 and MAE of 22.4, closely approaching Mercier's results (RMSE 31, MAE 19) using significantly less data. This demonstrates the capability of the model to generalise under extreme data constraints.

### 7.3. Cross-Season and Cross-Region Generalisation

To determine the generality of the frameworks and ensure they are not over-fitted to a single climate or camera configuration, we extend the evaluation in two stages. First, we test across multiple seasonal windows at our own site, training exclusively within each season and then testing on a full disjoint test year, to assess the advantage when irradiance dynamics and sun paths vary dramatically. Second, we apply our transformer-frameworks to a geographically distinct dataset. This stepped design allows for investigation of season-specific gains and cross-site generalisability, providing principled evaluation of the architectures transfer across regions and imaging conditions.

Table 7 – Performance Metrics Across Multiple Seasons

| Season (Training Window) | Model | MAE | RMSE | Seasonal Characteristics |
|---|---|---|---|---|
| S1 Spring 2016 (May – Jul '16) | Baseline | 28.80 | 40.33 | Brightness: μ=98.9 σ=12.9 |
| | GHI-Token | **21.42** | **31.07** | Kt: μ=0.42 σ=0.30 |
| | Dual Stream | 22.96 | 31.60 | Kd: μ=0.79 σ=0.29 |
| S2 Summer 2016 (Aug – Oct '16) | Baseline | 18.68 | 28.52 | Brightness: μ=98.2 σ=12.9 |
| | GHI-Token | 18.66 | 28.49 | Kt: μ=0.45 σ=0.61 |
| | Dual Stream | **16.56** | **25.21** | Kd: μ=0.74 σ=0.32 |
| S3 Autumn 2016 (Nov '16 – Jan '17) | Baseline | 31.06 | 51.62 | Brightness: μ=102.0 σ=13.1 |
| | GHI-Token | 28.61 | 52.87 | Kt: μ=0.43 σ=3.68 |
| | Dual Stream | **23.23** | **40.92** | Kd: μ=0.81 σ=0.27 |
| S4 Winter 2017 (Feb – Apr '17) | Baseline | 19.53 | 29.26 | Brightness: μ=101.1 σ=13.6 |
| | GHI-Token | 18.53 | 29.03 | Kt: μ=0.40 σ=0.53 |
| | Dual Stream | **17.00** | **26.15** | Kd: μ=0.79 σ=0.30 |

The Baseline yields the largest errors (MAE 18.7–31.1 W/m²; RMSE 28.5–51.6 W/m²) and its accuracy deteriorates with rising clearness-index variance, reaching a maximum in Autumn. Our frameworks improve accuracy in complementary ways: under the bright, largely homogeneous spring regime the GHI-Token design trims Baseline RMSE by 23 % (40.33 → 31.07 W/m²) and MAE by 26 %, whereas the Dual-Stream architecture dominates once clouds become thicker or more heterogeneous, delivering the lowest RMSE in summer, autumn and winter (25.21, 40.92 and 26.15 W/m²) and cutting Baseline errors by 12 %, 21 % and 11 %, respectively. Hence, while for either of the frameworks additional context is essential, a tokenised GHI prior suffices when irradiance patterns are smooth, whereas the dual-stream formulation is preferred whenever intra-season variability is moderate to high.

The next evaluation spans two distinct climates, temperate-oceanic (Cfb) at Chilbolton, UK and hot-summer Mediterranean (Csa) at Folsom, CA, USA providing a clear clearness/diffuse contrast that tests cross-climate generalisation.

Table 8 – Performance Metrics Across Multiple Locations

| Dataset | Variant | MAE | RMSE |
|---|---|---|---|
| Chilbolton | Baseline | 30.30 | 42.16 |
| Chilbolton | GHI-Token | **22.37** | **33.82** |
| Chilbolton | Dual Stream | 26.59 | 36.92 |
| Folsom | Baseline | 42.58 | 59.73 |
| Folsom | GHI-Token | **31.23** | **54.23** |
| Folsom | Dual Stream | 35.50 | 55.30 |

The cross-validation in Tables 7 and 8 confirms that the proposed data-efficient adaptations preserve their advantage under both seasonal and geographic domain shifts. When each model is trained in a single season and evaluated on a disjoint year (Table 7), the GHI-Token variant reduces Baseline RMSE by up to 23 % during the relatively homogeneous spring period, while the Dual-Stream architecture yields the lowest errors in the more heterogeneous summer, autumn, and winter windows (RMSE gains of 12–21 %). Extending the test to a climatically distinct site (Table 8) indicates that the same frameworks transferred to the Folsom dataset, still lower MAE by ≈ 27 % RMSE by ≈ 9 % relative to the baseline. Collectively, these results suggest that seasonal mismatches in cloud morphology and sun-path do not negate the benefit of our frameworks, and the generalisation across camera optics and temperate-to-sub-tropical climates, validating their suitability in data-scarce, multi-region scenarios.

## 8. Discussion

These results demonstrate the value of a multi-stream segmentation-based method that does not rely on training a fully supervised cloud classifier. By applying threshold-based segmentation, the proposed framework efficiently captures critical cloud-structure details without incurring substantial labelling overhead associated with supervised classification methods. This significantly enhances operational scalability, particularly valuable in remote or resource-constrained scenarios. Furthermore, the ability of the sequential pipeline to incorporate GHI as a guiding feature addresses the frequent shortage of direct DHI measurements in many regions.

In comparison to prior CNN- and transformer-based models trained on complete datasets, this hybrid framework maintains RMSE performance within 10 % of full-data baselines even when trained on only 1/12 of the available data, demonstrating substantial robustness under extreme data sparsity and effective utilisation of synthetic augmentation.

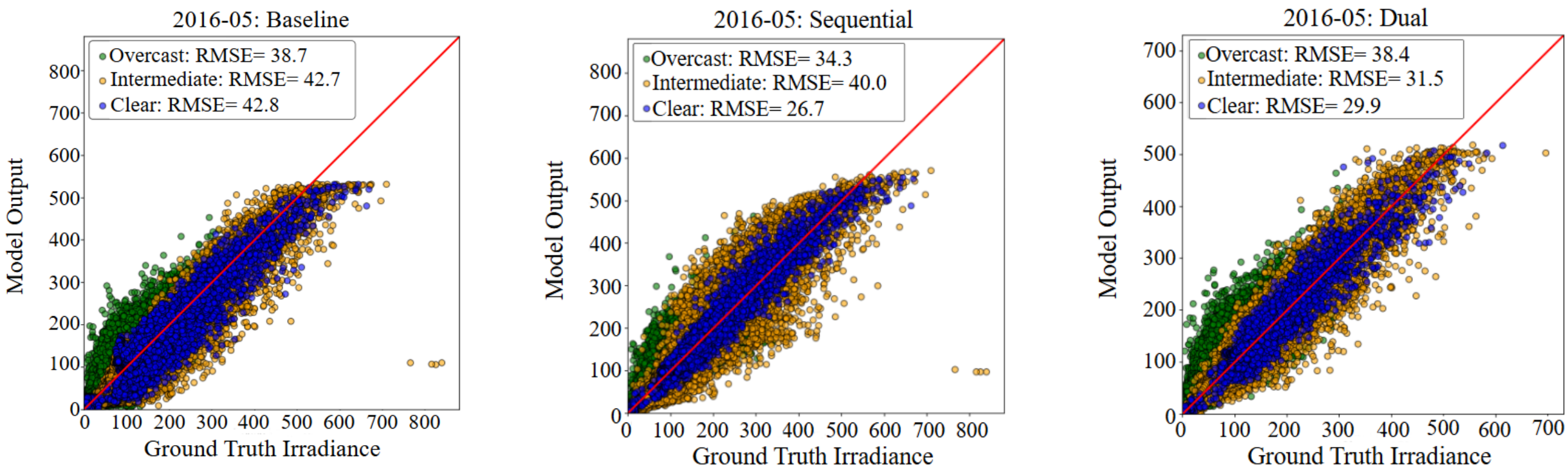

Figure 7 – Scatter plots comparing predicted versus ground truth DHI for three ViT models. (a) Baseline DeiT model, (b) Sequential DeiT model with GHI token injection, and (c) Dual-Parallel framework. Results are segmented by clearness index to illustrate model performance under varying sky conditions.

To evaluate model performance under varying sky conditions, the test set was stratified, as seen in Figure 7, by clearness index (CI) into overcast ($CI < 0.3$), intermediate ($0.3 \leq CI \leq 0.78$), and clear ($CI > 0.78$) regimes [45]. Results indicate that both the sequential with GHI injection and dual-stream approaches improve DHI prediction accuracy compared to the baseline. Notably, the GHI-Token injection method significantly reduces RMSE in clear conditions, while the dual-stream approach delivers its greatest improvement in intermediate and clear regimes. These enhancements contribute to more consistent irradiance estimates overall, although performance may vary slightly under different atmospheric conditions. These results underscore the capacity of the framework to generalise across diverse sky states without requiring cloud labels, supporting its suitability for deployment in data-limited environments.

Although the proposed framework demonstrates strong performance on data from CFARR, its generalisability to other locations or imaging systems may require fine-tuning, and minor threshold adjustments to the segmentation pipeline may be necessary to account for variations in camera hardware or atmospheric conditions. Future work should evaluate whether the observed generalisation across annual timescales extends to spatial domains and investigate the integration of temporal forecasting for short-term irradiance prediction. These aspects remain beyond the scope of the present study but represent valuable directions for future research.

## 9. Conclusion

Accurate DHI estimation is critical for PV optimisation and forecasting, yet ground sensors and years of sky images remain scarce. This study demonstrates that DeiTs can approach instrument-level DHI estimates without requiring long-term data archives. Two specialised architectures were devised: (i) a sequential DeiT that feeds a synthetic GHI token back into the network, and (ii) a parallel dual-stream DeiT that fuses raw and overlay-enhanced fisheye images. Both operate on calculable features, ETI and cyclic time, and lightweight segmentation thereby demonstrating that accurate DHI estimation is attainable when deterministic physical inputs are combined with lightweight cloud cues, in limited data scenarios.

Ten model variants were benchmarked on the CFARR data set (May 2016 – Aug 2018): Image-Only, Baseline, GHI-Appended, GHI-MLP, GHI-Token, Overlay-DeiT, Normal-DeiT, Parallel Dual-Stream, Concatenation-Normalised and Concatenation-Stream. Under a stringent one-month training / full-year test regime, the GHI-Token model cut the RMSE from 42.2 W/m² to 33.8 W/m² and improved $R^2$ from 0.77 to 0.84, while the dual-stream overlay reduced it to 36.9 W/m² and delivered the largest gains in partly cloudy conditions where cloud detail is critical. When the training window extended to a full year, both transformers converged to within 3 % of the Baseline RMSE and showed little advantage under uniformly overcast skies. These results delineate where the transformers excel (data-scarce, heterogeneous skies) and where they offer limited benefit (data-rich, solid overcast).

Both frameworks with 1/12 of the full dataset achieve RMSE values within 2-6 W/m² of a state-of-the-art ViT. Consequently, the proposed frameworks extract reliable DHI from a camera after only weeks of data, turning low-cost sky imagers into a scalable, low-maintenance measurement network for remote or under-instrumented regions and opening the door to camera-based forecasting, multi-site transfer learning and real-time power-system control.

## CRediT Authorship Contribution Statement

**David Hamlyn**: Writing – original draft, Visualisation, Software, Investigation, Conceptualisation. **Sunny Chaudhary:** Writing – review & editing, Software, Supervision, Conceptualisation. **Tasmiat Rahman:** Writing – review & editing, Supervision, Conceptualisation.

## Declaration of Competing Interest

The authors declare that they have no known competing financial interests or personal relationships that could have appeared to influence the work reported in this paper.

## Data Availability Statement

The underlying data and model code is available upon request.

## Acknowledgements

D. Hamlyn acknowledges the support of the Engineering and Physical Sciences Research Council (EPSRC) Doctoral Training Partnership (DTP) Funding. T. Rahman acknowledges support from EPSRC grant EP/X033333/1. We would like to thank the Chilbolton observatory for providing the data used in this publication.